\documentclass[acmsmall]{acmart}
\AtBeginDocument{%
  }

\usepackage{booktabs}
\usepackage{float}
\usepackage{soul}
\usepackage{hyperref}
\usepackage{multirow}
\usepackage{listings}
\usepackage{fancybox}
\usepackage{enumitem}
\usepackage{graphicx}
\usepackage{rotating}
\usepackage{tikz}
\usepackage{color}
\usepackage{array}
\usepackage{amsmath}
\usepackage{xspace}
\usepackage{url}
\usepackage{tikz}
\usepackage{caption}
\usepackage{pgfplots}
\usepackage{balance}
\usepackage{adjustbox}
\usepackage{subcaption}
\pgfplotsset{compat=1.16}
\usepackage{pgf-pie}
\usetikzlibrary{pgfplots.statistics,calc}
\usepackage{algorithm}
\usepackage{algorithmic}
\usepackage{adjustbox}
\usepackage{makecell}

\usepackage{tcolorbox}
\makeatletter
\newcommand{\mybox}[1]{%
	\setbox0=\hbox{#1}%
	\setlength{\@tempdima}{\dimexpr\wd0+13pt}%
	\begin{tcolorbox}[boxrule=0.5pt, colback=white, arc=4pt,
		left=6pt,right=6pt,top=6pt,bottom=6pt,boxsep=0pt]
		#1
	\end{tcolorbox}
}

\definecolor{songcolor}{RGB}{191,191,191}

\newcommand{\tool}{\textsc{CG$^{2}$}}

\begin{document}

\title{Better Specifications, Better Code: Fine-Tuning LLMs for Behavioral Contract Generation}


 \author{Brhane Teamrat Gidey}
 \affiliation{%
   \institution{York University}
   \department{Department of Electrical Engineering and Computer Science}
   \city{North York}
   \country{Canada}
 }
 \email{brhane.teamrat.gidey@gmail.com}

 \author{Tiantian Yang}
 \affiliation{%
   \institution{York University}
   \department{Department of Electrical Engineering and Computer Science}
   \city{North York}
   \country{Canada}
 }
 \email{yang11@my.yorku.ca}

 \author{Nikta Akbarpour}
 \affiliation{%
   \institution{York University}
   \department{Department of Electrical Engineering and Computer Science}
   \city{North York}
   \country{Canada}
 }
 \email{niktakbr@yorku.ca}

 \author{Simone Bartucci}
 \affiliation{%
   \institution{University of Calabria}
   \department{Computer Science}
   \city{Arcavacata}
   \country{Italy}
 }
 \email{simone.bartucci@unical.it}

 \author{Francesco Calimeri}
 \affiliation{%
   \institution{University of Calabria}
   \department{Computer Science}
   \city{Arcavacata}
   \country{Italy}
 }
 \email{francesco.calimeri@unical.it}

 \author{Song Wang}
 \orcid{0000-0003-0617-2877}
 \affiliation{%
   \institution{York University}
  \department{Department of Electrical Engineering and Computer Science}
   \city{North York}
   \country{Canada}
 }
 \email{wangsong@yorku.ca}


\begin{abstract}
 Large language models (LLMs) show strong code-generation capabilities but still struggle with functional correctness. Natural-language tasks often leave behavioral requirements, constraints, and corner cases implicit, forcing LLMs to infer both \emph{what} the program should do and \emph{how} to implement it simultaneously. Misinterpreting these implicit requirements can therefore propagate directly into functional errors. This motivates separating behavioral specification from implementation by first making the intended behavior explicit before generating code. 

We present {\tool}, a contract-guided code generation framework that uses behavioral contracts to bridge natural-language descriptions and code. {\tool} fine-tunes an LLM on pairs of task descriptions and contracts to generate structured behavioral specifications that explicitly capture program interfaces, preconditions, postconditions, and invariants, which then guide an unchanged downstream code-generation model. To mitigate errors in generated contracts, {\tool} further introduces an iterative semantic validation and repair process. Using the original task description as the semantic reference, {\tool} validates generated contracts to identify incorrect, vague, or missing clauses, as well as inconsistencies between machine-readable expressions and their natural-language descriptions. Identified defects are used as feedback to repair the contract, and repaired contracts are revalidated before being used for downstream code generation. 

We evaluate {\tool} on 1,007 held-out algorithmic code-generation tasks using five code LLMs: DeepSeek-Coder-V2-Lite-Instruct, Ministral-3-14B-Instruct-2512, Qwen2.5-Coder-14B-Instruct, Qwen2.5-Coder-32B-Instruct, and Qwen3.5-27B. Across all models, contract guidance consistently improves pass@1 over direct code generation, contract-specific fine-tuning provides further gains, and semantic validation and repair yield additional aggregate improvements. Overall, {\tool} improves pass@1 over direct generation by 1.49--3.27 percentage points, corresponding to relative improvements of 7.58\%--22.45\%, and over contract guidance without fine-tuning by 1.19--2.38 percentage points, corresponding to relative improvements of 7.04\%--14.65\%. Our manual contract-quality analysis further shows that contracts classified as semantically correct are consistently associated with stronger downstream code-generation performance than contracts containing semantic defects. These results provide evidence that explicitly translating natural-language requirements into validated behavioral contracts can improve functional correctness without modifying the downstream code-generation model.
\end{abstract}

\ccsdesc[500]{Software and its engineering~Software maintenance tools}
\ccsdesc[300]{Software and its engineering~Software verification and validation}

\keywords{Code Generation, LLMs, Program Specifications, Supervised Fine-Tuning}


\maketitle

\section{Introduction}
\label{sec:introduction}

Large language models (LLMs) have made substantial progress in code generation, with code-specialized models demonstrating strong performance across a wide range of programming benchmarks~\cite{roziere2023code,guo2024deepseek}. Nevertheless, functional correctness remains challenging for programming tasks with multiple interacting requirements, constraints, and corner cases~\cite{tian2025fixing,mu2024clarifygpt}.

Existing approaches attempt to reduce this problem by introducing intermediate reasoning before code generation. Chain-of-Thought~\cite{wei2022chain}, structured reasoning~\cite{li2025structured}, and self-planning approaches~\cite{jiang2024selfplanning} encourage models to reason about a task or construct an implementation plan before producing code. More specifically, approaches such as $\mu$FiX~\cite{tian2025fixing} explicitly improve an LLM's understanding of programming specifications through natural-language reasoning and test-case-driven feedback. These approaches demonstrate that separating requirement understanding from immediate implementation can improve code generation. 
However, the resulting understanding is typically represented as informal natural-language reasoning. Such representations can remain ambiguous, omit important behavioral requirements, or introduce assumptions that are not supported by the original task. Moreover, approaches that use execution feedback or execution-based verification~\cite{ni2023lever} assume that suitable test cases or other behavioral signals are available.

We explore a different direction: \textit{rather than representing requirement understanding only through informal reasoning, can an LLM explicitly translate a programming task into a structured behavioral specification \emph{before} generating its implementation?}

Behavioral contracts provide a natural representation for this purpose~\cite{meyer1992design}. A contract expresses program behavior
through explicit conditions and constraints, separating the specification
of expected behavior from the details of its implementation. Recent work
has demonstrated the potential of LLMs for synthesizing formal
specifications. SpecGen uses LLMs together with mutation and verification
to generate formal program specifications~\cite{ma2025specgen};
NL2ACSL translates natural-language requirements into ACSL function
contracts using retrieval and multiple agents~\cite{su2026nl2acsl}; and
VeriSpecGen decomposes natural-language requirements and iteratively
validates and repairs generated formal specifications~\cite{ye2026verispecgen}.
These studies demonstrate that LLMs can recover useful specifications from
natural language or code. However, specification generation has primarily
been studied as an end goal for program analysis and verification, rather than as an active intermediate reasoning step to guide LLMs. Furthermore, relying on zero-shot prompting to generate these intermediate contracts is risky. As we observe in our experiments, generated contracts can introduce unsupported constraints or misrepresent task requirements, producing flawed specifications that may mislead downstream implementation. More generally, prior work has highlighted limitations in the ability of LLMs to reliably identify and correct their own reasoning errors~\cite{huang2024large}. Therefore, how to systematically generate, validate, and specialize intermediate contracts to reliably improve functional correctness remains a critical open challenge.

\begin{figure*}[t]
    \centering
    \includegraphics[width=\textwidth]{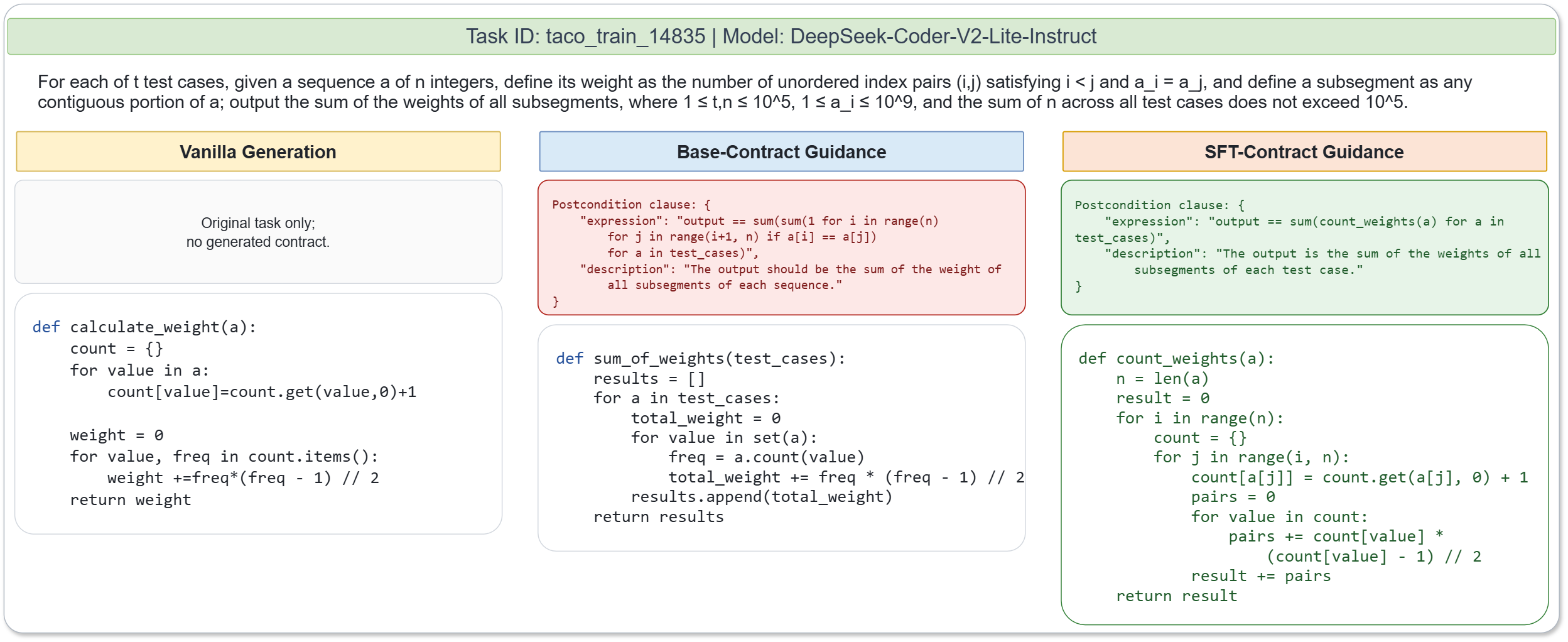}
    \caption{Motivating example from the held-out TACO evaluation set
    (task ID: \texttt{taco\_train\_14835}). The Vanilla implementation fails, while the Base-Contract-guided implementation fails due to a semantic mismatch. The SFT-Contract-guided implementation succeeds with a semantically consistent contract.}
    \label{fig:motivation}
\end{figure*}

Figure~\ref{fig:motivation} illustrates this challenge using a real example, \texttt{taco\_train\_14835}, from the TACO algorithmic code-generation dataset~\cite{li2023taco}. To ensure a controlled comparison, we use the same model, DeepSeek-Coder-V2-Lite-Instruct~\cite{zhu2024deepseek}, as the downstream code generator across all settings, with the natural-language task description as input. The three settings differ only in how the contract is introduced. In \textit{Vanilla}, the model generates code directly from the task description without any contract. In \textit{Base Contract}, the original DeepSeek-Coder-V2-Lite-Instruct first generates a behavioral contract from the task description, which is then provided together with the task to guide code generation. In \textit{SFT-Contract}, the contract is instead generated by DeepSeek-Coder-V2-Lite-Instruct fine-tuned using our contract-specific approach, while the downstream code generator remains the original, unchanged model. This controlled setup allows us to isolate the effect of introducing behavioral contracts and, more importantly, improving their quality through contract-specific fine-tuning.

As shown in Figure~\ref{fig:motivation}, the quality of the intermediate contract can affect the generated solution. In the \textit{Vanilla} setting, the model fails to correctly capture the required behavior and produces an incorrect implementation. Introducing a \textit{Base Contract} makes the behavioral requirements more explicit; however, the generated contract contains a semantic mismatch: although its natural-language description correctly describes the required aggregation over all contiguous subsegments, the corresponding machine-readable expression computes the metric only once. This incorrect specification propagates to the downstream code, resulting in another failed implementation. In contrast, the \textit{SFT Contract} correctly and consistently captures the required behavior in both its natural-language description and machine-readable expression. Guided by this higher-quality contract, the same downstream model generates an implementation that passes the test suite. This example highlights that simply introducing a contract is not sufficient; the semantic quality and consistency of the generated contract are critical to its effectiveness in guiding code generation. 


Motivated by these observations, we present
\textbf{{\tool}: Contract-Guided LLM Code Generation}, a framework that introduces automatically generated behavioral contracts as intermediate representations between natural-language programming tasks and their implementations. Rather than adapting the downstream code-generation model, CG$^{2}$ specializes the intermediate contract-generation step through supervised fine-tuning (SFT). To construct the contract-specific SFT dataset, we first use GPT-4o-mini~\cite{openai2024gpt4osystemcard} exclusively for training-data construction to generate candidate contracts, which are subsequently filtered, reviewed, and cleaned by the authors. Because even fine-tuned models may generate contracts that omit or misrepresent task requirements, CG$^{2}$ further performs semantic validation and repair before using the contracts for code generation. 
Specifically, generated contracts are checked against the original programming task, identified semantic defects are used to guide contract repair, and repaired contracts are revalidated before being accepted. Throughout this process, the original task serves as the primary semantic reference, while the downstream code-generation model remains unchanged. 


We evaluate CG$^{2}$ on a fixed set of 1,007 held-out programming tasks from TACO, a large-scale competitive-programming benchmark with executable test cases and difficulty annotations~\cite{li2023taco}.

Our evaluation covers five code LLMs under four controlled conditions:
\textit{Vanilla}, \textit{Base Contract}, \textit{SFT}, and
\textit{Validated SFT}. The downstream code-generation model and settings
remain fixed while only contract generation and refinement vary.
Section~\ref{sec:experimental-setup} provides the complete methodology.

Across all five models, pass@1 increases from \textit{Vanilla} to \textit{Base Contract}, \textit{SFT}, and \textit{Validated SFT}. Overall, CG$^{2}$ improves pass@1 over direct generation by 1.49--3.27 percentage points, corresponding to relative improvements of 7.58\%--22.45\%, and over contract guidance without fine-tuning by 1.19--2.38 percentage points, corresponding to relative improvements of 7.04\%--14.65\%. 
This paper makes the following contributions:
\begin{itemize}




   \item \textbf{Behavioral contracts as an intermediate representation for code generation.}
    We formulate contract-guided code generation, in which an LLM first translates a natural-language programming task into a structured behavioral contract and then uses the contract to guide implementation. This formulation explicitly separates reasoning about \emph{what} behavior a program should satisfy from \emph{how} that behavior should be implemented.

    \item \textbf{Learning to generate contracts for downstream code generation.}
    We develop a contract-specific supervised fine-tuning approach that specializes LLMs to synthesize structured contracts. Unlike approaches that adapt the code generator itself, our approach improves the intermediate specification while keeping the downstream code-generating model unchanged.

    \item \textbf{Semantic quality control for generated contracts.}
    We introduce an iterative validation and repair procedure that checks synthesized contracts against the original programming task, identifies semantic defects, and repairs them before code generation, without relying on reference implementations, generated code, or execution feedback.

\item \textbf{Large-scale empirical analysis of contract-guided code
generation and contract quality.}
We evaluate {\tool} on 1,007 held-out TACO tasks across five LLMs under four
controlled conditions. Beyond overall code-generation performance, we manually analyze the semantic
quality of generated contracts, examine the relationship between
contract-quality categories and downstream pass@1, and study the contribution
of individual contract components.

\end{itemize}

The remainder of this paper is organized as follows. Section~\ref{sec:background} reviews the related work. Section~\ref{sec:approach} presents CG$^{2}$. Section~\ref{sec:experimental-setup} describes the research questions and experimental methodology. Section~\ref{sec:results} reports the empirical results. Section~\ref{sec:discussion} discusses the key findings. Section~\ref{sec:threats} presents the threats to validity, and Section~\ref{sec:conclusion} concludes the paper.
\section{Background and Related Work}
\label{sec:background}

\subsection{Program Specifications and Specification-Guided Code Generation}

Program specifications describe the expected behavior of a program without
prescribing a particular implementation strategy. Design by Contract characterizes program behavior through preconditions, postconditions, and
invariants~\cite{meyer1992design}. Preconditions specify conditions that
must hold before execution, postconditions describe properties that must hold after execution, and invariants express properties that must remain true at designated points during execution. Verification-oriented languages such as
Dafny provide machine-checkable constructs for expressing and checking such behavioral properties~\cite{koenig2012dafny}.

Prior work shows that explicitly supplied specifications can improve
code-generation accuracy and constrain generated behavior. Newcomb et al.~\cite{newcomb2025contracts} show that adding preconditions and
postconditions to class-level code-generation prompts can improve initial
generation accuracy, with particularly pronounced gains for models with
fewer parameters or weaker baseline performance. ContractEval~\cite{lim2026contracteval} further evaluates whether generated
code satisfies explicit behavioral assertions, highlighting the importance of
contract satisfaction when assessing code-generation systems. Chakraborty et al.~\cite{chakraborty2025neural} investigate the synthesis
of F$^{*}$ programs and proofs from formal specifications expressed as
F$^{*}$ types. Their results show that fine-tuned smaller models can perform competitively with substantially larger prompted models on these specification-guided synthesis
tasks.

Although these studies demonstrate the benefits of specification guidance,
the guiding specification is supplied to the model rather than first inferred
from the natural-language programming task. In {\tool}, the behavioral
contract must itself be synthesized before it can guide code generation.
Errors or omissions in the generated contract may therefore misrepresent the
intended behavior and provide misleading guidance to the downstream
implementation. This makes the semantic quality of the generated
specification an additional concern.

\subsection{LLM-Driven Specification Synthesis, Validation, and Refinement}

Recent studies use LLMs to synthesize and refine formal specifications. AutoSpec~\cite{wen2024autospec} generates candidate specifications for C programs and iteratively validates them using static analysis and program verification, while SpecGen~\cite{ma2025specgen} combines specification generation with mutation and verification-based refinement. DbC-GPT~\cite{leite2026dbcgpt} similarly generates and verifies postconditions for Solidity programs. These approaches rely on an existing implementation as a reference for assessing generated specifications. 
Other work derives specifications directly from natural-language requirements. Endres et al.~\cite{endres2024postconditions} generate formal postconditions from documentation, while NL2Contract~\cite{richter2025beyond} generates preconditions and postconditions and evaluates their soundness and verification utility. NL2ACSL~\cite{su2026nl2acsl} uses specialized agents to translate natural-language requirements into ACSL contracts, and related work highlights the difficulty of ensuring that formalized specifications faithfully capture informal intent~\cite{lahiri2024intent}. Other approaches refine generated outputs through feedback or external validation, including Self-Debug~\cite{chen2024selfdebug}, Self-Refine~\cite{madaan2023selfrefine}, 3DGen~\cite{fakhoury2025_3dgen}, and Clover~\cite{sun2024clover}.

Prior work shows that behavioral specifications can guide code generation and that LLMs can synthesize specifications from programs or natural-language requirements. {\tool} connects these directions by generating behavioral contracts from programming tasks and using them to guide downstream code generation. To mitigate contract errors, {\tool} validates and repairs contracts against the original task without using reference implementations, generated code, or execution outcomes.
\section{Approach}
\label{sec:approach}
\subsection{Framework Overview}

\begin{figure}[t!]
    \centering
    \includegraphics[width=\columnwidth]{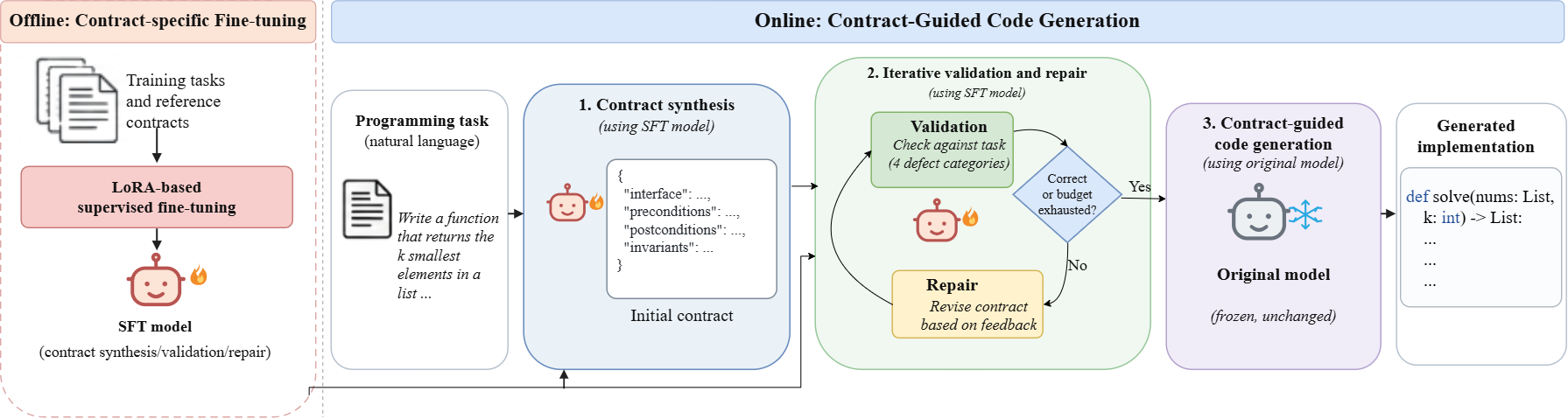}
    
    \caption{Overview of {\tool}.}
    \label{fig:overview}
\end{figure}

Figure~\ref{fig:overview} shows the overview of {\tool}. Given a natural-language programming task, we first synthesize a structured contract using a contract-specialized model obtained via LoRA-based supervised fine-tuning. We then iteratively validate and repair the generated contract to
identify semantic defects and attempt to correct them. Finally, the selected contract, along with the original task, guides a frozen, off-the-shelf code-generation model to produce the implementation.


\subsection{Contract-specific Fine-tuning}

During offline fine-tuning, we specialize LLMs for behavioral contract synthesis through supervised fine-tuning (SFT) using Low-Rank Adaptation (LoRA)~\cite{hu2021lora}. Rather than fine-tuning the model to generate code directly, our goal is to improve its ability to translate natural-language programming tasks into explicit behavioral specifications. LoRA keeps the pretrained model weights frozen and learns low-rank updates to selected weight matrices, enabling parameter-efficient specialization.
Each training example pairs a natural-language programming task with a target behavioral contract. We formulate each example as a conversation consisting of a system instruction defining the contract-generation task, a user message containing the programming task, and an assistant response containing the target contract. The model is trained to generate structured contracts that capture four key elements of program behavior: the \textit{interface}, \textit{preconditions}, \textit{postconditions}, and \textit{invariants}.

We refer to the original model with the trained LoRA adapter enabled as the \textit{SFT model}. The SFT model is used for contract synthesis as well as subsequent contract validation and repair. Importantly, the downstream code-generation stage uses the original model checkpoint without the LoRA adapter. This design isolates the adaptation to contract generation and refinement, allowing us to study whether improving the intermediate behavioral specification can improve code generation without modifying the code-generating model itself. Section~\ref{sec:4.3} describes the construction of the SFT corpus, and Section~\ref{sec:4.4} details the fine-tuning configuration.


\subsection{Contract-guided Code Generation}

\subsubsection{Contract Synthesis}

Contracts are represented as structured JSON objects that explicitly capture four aspects of program behavior: the \textit{interface}, \textit{preconditions}, \textit{postconditions}, and \textit{invariants}. The interface defines how the program is invoked and specifies its expected inputs and outputs. Preconditions capture conditions that must hold for valid inputs, postconditions specify properties that a correct output must satisfy, and invariants describe properties that should remain valid throughout relevant parts of the computation. Each contract may contain multiple preconditions, postconditions, and invariants, with each clause consisting of a unique identifier, a machine-readable expression, and a corresponding natural-language description. 

During contract synthesis, the SFT model receives the complete natural-language programming task and generates a structured contract before any code is produced. The model is instructed to faithfully translate the behavioral requirements of the original task into explicit contract clauses, including relevant constraints and corner cases. Importantly, the generated contract should make implicit requirements explicit while avoiding new assumptions or constraints that are not supported by the original task.

\subsubsection{Iterative Validation and Repair of Contracts}
\label{sec:contract-validation}

Because automatically generated contracts may omit or misrepresent
task requirements, we apply a validation and repair step to the
contracts generated by the SFT model. The validation and repair
process is independent of program execution and does not use a
reference implementation, generated code, or execution outcomes.
Instead, the original programming task serves as the primary
semantic reference for evaluating and repairing each contract.
During validation, the model audits the interface, preconditions,
postconditions, and invariants for consistency with the original
task. We define four semantic defect categories for contract
validation:

\begin{itemize}
    \item \texttt{incorrect\_clause} indicates that a clause
    contradicts the task description, states an incorrect condition,
    introduces an unsupported restriction, or misrepresents the
    required interface.

    \item \texttt{vague\_clause} indicates that a clause does not
    provide sufficiently precise semantic information, such as when
    it refers to a ``correct result'' without defining what it means
    for the result to be correct.

    \item \texttt{missing\_clause} indicates that an important
    requirement from the original task is not represented in the
    contract.

    \item \texttt{expression\_description\_mismatch} indicates that
    the machine-readable expression and its natural-language
    description specify different requirements.
\end{itemize}

The verifier follows an expression-led policy: expressions are the
primary semantic target, while descriptions are used to interpret
their meaning and identify material contradictions. Prose-only
concerns do not normally trigger repair when the expression already
captures the task requirement. Proposed semantic defects undergo a
separate confirmation step before guiding repair.
The semantic defect categories characterize the type of issue identified in
the contract, whereas the validator verdict summarizes whether an actionable
semantic defect has been established.
The validator returns \texttt{CORRECT}, \texttt{DEFECTIVE}, or
\texttt{UNCERTAIN}. \texttt{CORRECT} indicates that no actionable
defect was established during the validation procedure, while
\texttt{DEFECTIVE} indicates that a substantive defect was
established. \texttt{UNCERTAIN} indicates that the available
assessment does not support a confident decision. Contracts judged
\texttt{CORRECT} are preserved. Contracts judged \texttt{DEFECTIVE}
undergo repair, while \texttt{UNCERTAIN} contracts are normally
preserved unless an actionable issue justifies a repair attempt.

When repair is required, the SFT model receives the original task,
the current contract, and the identified defects. The model is asked
to produce a revised contract that addresses the flagged defects
while preserving content that is already correct. Since a repair
may introduce additional errors, the repaired contract goes through
a deterministic structural check before the validation step is
applied again. Before a repaired candidate replaces the original contract, the proposed
changes undergo an additional semantic review against the original programming
task. The candidate is accepted only if this review identifies no actionable
semantic defect.

The procedure terminates when the contract is preserved without repair, a repaired candidate is accepted, or the repair budget is exhausted (up to five repair attempts). A repaired candidate replaces the original contract only after passing both structural and semantic validation. If no candidate is accepted within five attempts, the original contract is preserved. 

\subsubsection{Contract-guided Code Generation}

The original programming task and a generated contract are provided
to the downstream model for code generation. The contract is produced
by either the original model or the SFT model and may subsequently
undergo validation and repair. The code-generation stage uses the
original checkpoint without the contract-specific LoRA adapter.
Thus, the same stage supports guidance using base-generated
contracts, SFT-generated contracts, and contracts selected after
validation and repair. Section~\ref{sec:experimental-setup} defines
these experimental conditions.

The model is instructed to interpret the task and contract together,
using the contract as structured guidance for making behavioral
requirements explicit. The generated implementation is expected to
satisfy the contract's task-consistent preconditions, postconditions,
and invariants while preserving the required interface and
input-output behavior. The original task remains authoritative when
the contract is incomplete, ambiguous, or inconsistent with its
requirements. 
\section{Experimental Setup}
\label{sec:experimental-setup}

\subsection{Research Questions}

We evaluate whether contract guidance, contract-specific fine-tuning, and
post-fine-tuning contract validation improve the functional correctness of
generated code. Our study addresses the following research questions:

\begin{itemize}

    \item \textbf{RQ1: Does contract guidance improve the functional
    correctness of LLM-generated code (Effectiveness of Contract Guidance)?}
    We compare \textit{Vanilla} with \textit{Base Contract}. In \textit{Vanilla}, the
    original model generates code directly from the programming task, whereas
    in \textit{Base Contract}, the same model first generates a behavioral contract and
    then uses the task together with the contract to generate code. This
    comparison isolates the effect of introducing model-generated contract
    guidance without contract-specific fine-tuning.

    \item \textbf{RQ2: Does specializing an LLM for contract synthesis
    improve downstream code correctness (Effectiveness of Contract-Specific
    Fine-Tuning)?}
    We compare \textit{Base Contract} with \textit{SFT}. \textit{Base Contract} uses contracts
    generated by the original model, whereas \textit{SFT} uses contracts
    generated by the contract-specific SFT model. In both conditions, the
    final code is generated by the same original checkpoint without the
    contract-specific adapter. This comparison isolates the effect of
    contract-specific fine-tuning on downstream code-generation performance.

    \item \textbf{RQ3: Does semantic validation and repair of contracts
    generated by the fine-tuned model provide additional improvements in
    downstream code correctness (Contract Validation)?}
    We compare \textit{SFT} with \textit{Validated SFT}. The former directly
    uses contracts generated by the SFT model, whereas the latter validates
    and, when necessary, repairs these contracts before code generation.
    In addition to overall pass@1, we analyze task-level transitions from
    failures to passes and from passes to failures. This comparison evaluates
    both the improvements and regressions associated with contract validation
    and repair.

    \item \textbf{RQ4: How is the semantic quality of behavioral contracts
    associated with downstream code-generation performance
    (Effect of Contract Quality)?}
    We categorize contracts as semantically correct or as containing different
    types of semantic defects, including incorrect clauses, vague clauses,
    missing clauses, and expression--description mismatches. We then compare the downstream pass@1 associated with each category to examine
the association between contract semantic quality and code-generation
correctness.

\end{itemize}

\subsection{Experimental Design}

To answer our research questions, we evaluate {\tool} under four controlled
experimental conditions. Across all conditions, the final code implementation
is generated using the original provider-released model checkpoint without
contract-specific fine-tuning. We vary only how the behavioral contract is
generated and refined, allowing us to isolate the effects of contract guidance,
contract-specific fine-tuning, and semantic validation and repair.

The four experimental conditions are as follows:

\begin{enumerate}

    \item \textbf{Vanilla.}
    The original model generates code directly from the programming task
    without contract guidance.

    \item \textbf{Base Contract.}
    The original model first generates a behavioral contract from the
    programming task. The task and generated contract are then provided to the
    same original model for code generation.

    \item \textbf{SFT.}
    The contract-specific SFT model generates a behavioral contract from the
    programming task. The task and generated contract are then provided to the
    original model, without the contract-specific adapter, for code generation.

    \item \textbf{Validated SFT.}
    The contract-specific SFT model first generates a behavioral contract,
    which is then subjected to the semantic validation and repair procedure
    described in Section~\ref{sec:contract-validation}. The task and final
    selected contract are subsequently provided to the original model,
    without the contract-specific adapter, for code generation.

\end{enumerate}

\begin{figure}[t]
    \centering
    \includegraphics[width=0.9\columnwidth]{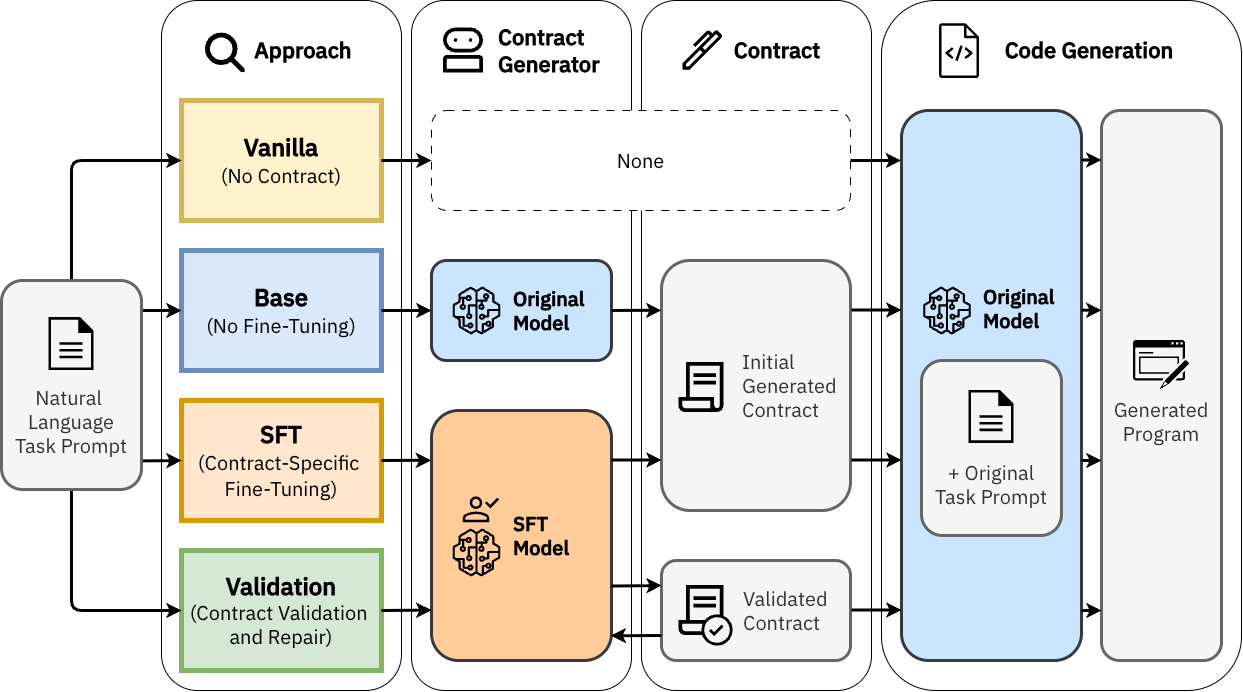}
    \caption{Overview of the four experimental conditions.}
    \label{fig:experimental-conditions}
\end{figure}

Figure~\ref{fig:experimental-conditions} summarizes the four experimental
conditions. Comparing \textit{Vanilla} with \textit{Base Contract} isolates the effect
of introducing a model-generated behavioral contract (RQ1). Comparing
\textit{Base Contract} with \textit{SFT} isolates the effect of contract-specific
fine-tuning while keeping the downstream code-generation model unchanged
(RQ2). Comparing \textit{SFT} with \textit{Validated SFT} evaluates the
additional effect of semantic contract validation and repair (RQ3). Finally,
we analyze the final contracts used in the \textit{Validated SFT} condition
using manually assigned semantic categories to characterize contract quality
and examine its association with downstream code-generation performance
(RQ4).

\subsection{Dataset Construction}
\label{sec:4.3}

We construct the fine-tuning and evaluation datasets from TACO, a large-scale
competitive-programming dataset containing executable test cases, difficulty
labels, and topic annotations~\cite{li2023taco}. We first select tasks from
four algorithmic categories: \textit{String}, \textit{Constructive},
\textit{Graph}, and \textit{Tree}. After deduplication, we exclude tasks
without usable test cases, resulting in 5,593 unique tasks. We further remove
tasks containing fewer than three test cases, because tasks with very limited
test coverage provide insufficient evidence for reliably assessing the
functional correctness of generated programs. This filtering leaves
4,721 tasks.

We assign 3,714 of the 4,721 tasks to the fine-tuning data-construction pool
and the remaining 1,007 tasks to the held-out evaluation set. To construct the
fine-tuning corpus, we use GPT-4o-mini to generate a behavioral contract for
each of the 3,714 construction tasks. For each task, the generated contract
and original task description are then provided to GPT-4o-mini to generate a
program, which is executed against the available TACO test cases. We retain
task--contract pairs whose generated programs pass at least one test case as
candidates for further review. Passing at least one test is used only as a
coarse automatic screening criterion and is not treated as evidence that the
corresponding contract is semantically correct.

The retained candidates subsequently undergo manual review, cleaning, and
deduplication to form the final supervised fine-tuning corpus. Specifically,
two authors independently review each candidate task--contract pair against
the original programming task. The reviewers assess whether the generated
contract faithfully captures the task requirements, including the interface,
preconditions, postconditions, and invariants, and check for incorrect or
unsupported constraints, missing requirements, ambiguous clauses, and
inconsistencies between machine-readable expressions and their corresponding
natural-language descriptions. Candidates containing substantial semantic
errors are removed, while minor formatting and structural inconsistencies are
cleaned to maintain a consistent contract representation across the corpus.

To assess the reliability of this manual review, we compute Cohen's $\kappa$
over the two reviewers' independent retain/remove judgments for candidate
task--contract pairs, obtaining a value of 0.85, indicating high
inter-reviewer agreement. All disagreements are
subsequently discussed among the authors until consensus is reached. Finally,
we remove duplicate and near-duplicate task--contract pairs to reduce
redundancy and potential data leakage.

The resulting corpus contains 2,165 task--contract examples, of which 2,056
are used for training and 109 for validation. The final held-out evaluation
set contains 1,007 tasks and is used exclusively for evaluating the four
experimental conditions.

\subsection{Models and Experimental Configuration}
\label{sec:4.4}
We evaluate five code-oriented instruction-tuned LLMs from three model
families: DeepSeek-Coder-V2-Lite-Instruct~\cite{zhu2024deepseek},
Ministral-3-14B-Instruct-2512~\cite{liu2026ministral3},
Qwen2.5-Coder-14B-Instruct~\cite{hui2024qwen25coder},
Qwen2.5-Coder-32B-Instruct~\cite{hui2024qwen25coder}, and
Qwen3.5-27B~\cite{qwen2026qwen35}. We focus on models that can be fine-tuned
using our available computational resources.
To avoid model-specific hyperparameter tuning, we use the same LoRA
hyperparameters and optimization settings across all five models. Each model is fine-tuned for four epochs
using LoRA with rank $r=8$, scaling factor $\alpha=16$, and dropout $0.05$.
For Qwen2.5-Coder-32B-Instruct and Qwen3.5-27B, we use QLoRA-style
4-bit NF4 quantization during fine-tuning to reduce GPU memory usage
~\cite{dettmers2023qlora}, while retaining the same LoRA hyperparameters.
We use AdamW with an initial learning rate of $2\times10^{-5}$, cosine
learning-rate decay, a warmup ratio of $0.03$, weight decay of $0.01$, and a
maximum gradient norm of $1.0$. The per-device batch size is 1 with
16 gradient-accumulation steps, yielding an effective batch size of 16.
The maximum training sequence length is 4,096 tokens, training uses BF16
precision, and the random seed is fixed to 42. These hyperparameters are held
fixed across models and are not tuned on the held-out evaluation set.
Table~\ref{tab:configuration} summarizes the configuration.

All inference conditions use deterministic greedy decoding with sampling
disabled and one candidate per task. Contract generation receives the complete
programming task and known interface metadata, with a maximum output length of
1,536 tokens. Code generation receives either the complete task alone or the
task together with the selected contract, with a maximum output length of
4,096 tokens.

We conduct all experiments on a Dell PowerEdge R760xa (2U) server equipped
with two 32-core Intel Xeon Gold processors and four NVIDIA GeForce RTX 4090
GPUs, each with 24\,GB of GPU memory (96\,GB in total). All fine-tuning and
inference experiments are performed on the same computing infrastructure.

\begin{table}[t]
\centering
\caption{Common fine-tuning and inference configuration.}
\label{tab:configuration}
\begin{tabular}{ll@{\hspace{3em}}ll}
\hline
\textbf{Parameter} & \textbf{Value} &
\textbf{Parameter} & \textbf{Value} \\
\hline
LoRA rank & 8 &
Effective batch size & 16 \\
LoRA $\alpha$ & 16 &
Maximum training length & 4,096 tokens \\
LoRA dropout & 0.05 &
Contract generation limit & 1,536 tokens \\
Optimizer & AdamW &
Code generation limit & 4,096 tokens \\
Learning rate & $2\times10^{-5}$ &
Precision & BF16 \\
Scheduler & Cosine &
Decoding & Greedy \\
Warmup ratio & 0.03 &
Candidates per task & 1 \\
Weight decay & 0.01 &
Random seed & 42 \\
\hline
\end{tabular}
\end{table}

\subsection{Evaluation and Analysis}

Across all experimental conditions, generated programs are evaluated using
the local TACO execution harness. A generated implementation is considered
correct only if it passes the entire associated test suite. Tests are executed
sequentially with a 60-second timeout per test, and evaluation stops for a task
at the first syntax, runtime, timeout, or logical failure.

For each experimental condition, exactly one program is generated for each of
the 1,007 evaluation tasks. Consequently, pass@1 is measured directly as

\begin{equation}
    \text{pass@1} =
    \frac{\text{number of tasks passed}}
         {\text{number of evaluated tasks}}.
\end{equation}

We do not use retries, best-of-$N$ selection, or multiple-sample pass@$k$
estimation. Our goal is to measure the direct effect of contract guidance on a
single deterministic code-generation attempt. Allowing multiple generations
or selecting among multiple candidates could introduce additional sampling
and selection effects, making it more difficult to isolate the contribution
of the contract. Therefore, each model generates exactly one solution per task
under deterministic greedy decoding, and we report pass@1 as the primary
measure of functional correctness.
\section{Result Analysis}
\label{sec:results}

Table~\ref{tab:overall_results} summarizes performance on the
1,007 held-out TACO tasks under the four experimental conditions
defined in Section~\ref{sec:experimental-setup}. We organize the
analysis by research question. All gains refer to net changes
in passed-task counts, and percentage-point differences are
calculated from the underlying counts before rounding.

\begin{table}[t!]
    \centering
    \caption{Results on 1,007 held-out TACO tasks.
    Each entry reports the number of passed tasks and pass@1.}
    \label{tab:overall_results}
    \small
    \setlength{\tabcolsep}{4pt}
    \renewcommand{\arraystretch}{1.2}
    \begin{tabular*}{\linewidth}{
        @{\extracolsep{\fill}}lcccc@{}
    }
        \toprule
        \textbf{Model}
        & \textbf{Vanilla}
        & \textbf{Base Contract}
        & \textbf{SFT}
        & \textbf{Validated SFT} \\
        \midrule
        DeepSeek-Coder-V2-Lite-Instruct
        & 147 (14.60\%)
        & 157 (15.59\%)
        & 168 (16.68\%)
        & 180 (17.87\%) \\

        Ministral-3-14B-Instruct-2512
        & 181 (17.97\%)
        & 188 (18.67\%)
        & 206 (20.46\%)
        & 212 (21.05\%) \\

        Qwen2.5-Coder-14B-Instruct
        & 157 (15.59\%)
        & 160 (15.89\%)
        & 167 (16.58\%)
        & 178 (17.68\%) \\

        Qwen2.5-Coder-32B-Instruct
        & 159 (15.79\%)
        & 163 (16.19\%)
        & 171 (16.98\%)
        & 175 (17.38\%) \\

        Qwen3.5-27B
        & 198 (19.66\%)
        & 199 (19.76\%)
        & 207 (20.56\%)
        & 213 (21.15\%) \\
        \bottomrule
    \end{tabular*}
\end{table}

\subsection{RQ1: Effectiveness of Contract Guidance}

\paragraph{Approach.}
To investigate whether introducing a behavioral contract improves the functional correctness of LLM-generated code, we compare the \textit{Vanilla} and \textit{Base Contract} conditions. 
In the \textit{Vanilla} condition, the model receives only the original programming task and directly generates a code solution. In the \textit{Base Contract} condition, the same original model first generates a behavioral contract from the programming task. The generated contract is then provided together with the original task to the same model for code generation. Thus, the only difference between the two conditions is the presence of the model-generated behavioral contract during code generation. We evaluate both conditions on the same 1,007 held-out tasks and compare their
pass@1, where a generated program is considered correct only if it passes all
available test cases for the corresponding task.

\paragraph{Results.}
Table~\ref{tab:overall_results} reports the results. Introducing a behavioral contract improves pass@1 for all five evaluated models, although the magnitude of the improvement varies across model families. For DeepSeek-Coder-V2-Lite-Instruct, pass@1 increases from 14.60\% (147/1,007) under \textit{Vanilla} to 15.59\% (157/1,007) under \textit{Base Contract}, corresponding to 10 additional correctly solved tasks and an absolute improvement of 0.99 percentage points. Ministral-3-14B-Instruct-2512 improves from 17.97\% (181/1,007) to 18.67\% (188/1,007), gaining seven solved tasks and 0.70 percentage points.

Similar, but smaller, improvements are observed for the Qwen models. Qwen2.5-Coder-14B-Instruct increases from 15.59\% (157/1,007) to 15.89\% (160/1,007), corresponding to three additional solved tasks and a 0.30-percentage-point improvement. Qwen2.5-Coder-32B-Instruct improves from 15.79\% (159/1,007) to 16.19\% (163/1,007), gaining four tasks and 0.40 percentage points. Qwen3.5-27B exhibits the smallest change, increasing from 19.66\% (198/1,007) to 19.76\% (199/1,007), corresponding to one additional solved task and an improvement of 0.10 percentage points.

Overall, \textit{Base Contract} outperforms \textit{Vanilla} for every evaluated model, with absolute improvements ranging from 0.10 to 0.99 percentage points. The consistent direction of the changes suggests that making behavioral requirements explicit through an intermediate contract can provide useful guidance for downstream code generation. However, the gains from contracts generated by the original models are relatively modest, indicating that simply introducing a behavioral contract is not sufficient to produce substantial improvements. The quality of the generated contract may play an important role in determining the effectiveness of contract guidance. 

\mybox{\textbf{Answer to RQ1:}
Introducing model-generated behavioral contracts consistently improves pass@1 across all five evaluated models. However, the improvements are modest, motivating the need to improve the quality of the generated contracts.
}

\subsection{RQ2: Effectiveness of Contract-Specific Fine-Tuning}

\paragraph{Approach.}
To investigate whether specializing an LLM for behavioral contract synthesis improves downstream code generation, we compare the \textit{Base Contract} and \textit{SFT} conditions. 
In the \textit{Base Contract} condition, the original provider-released model generates a behavioral contract from the programming task. In the \textit{SFT} condition, the same base model equipped with our contract-specific LoRA adapter generates the contract. Importantly, after contract generation, both conditions use the same original provider-released checkpoint, without the LoRA adapter, to generate the final code implementation. The downstream model receives the same original programming task together with the corresponding generated contract. Therefore, the primary difference between the two conditions is whether the behavioral contract is generated by the original model or by the model specialized for contract synthesis. We evaluate both conditions on the same 1,007 held-out tasks and compare their pass@1 results.

\paragraph{Results.}
As shown in Table~\ref{tab:overall_results}, contract-specific fine-tuning consistently improves downstream pass@1 across all five evaluated models. For DeepSeek-Coder-V2-Lite-Instruct, pass@1 increases from 15.59\% (157/1,007) under \textit{Base Contract} to 16.68\% (168/1,007) under \textit{SFT}, corresponding to 11 additional correctly solved tasks and an absolute improvement of 1.09 percentage points. Ministral-3-14B-Instruct-2512 exhibits the largest improvement, increasing from 18.67\% (188/1,007) to 20.46\% (206/1,007), with 18 additional solved tasks and a gain of 1.79 percentage points.

The Qwen models exhibit a similar pattern. Qwen2.5-Coder-14B-Instruct improves from 15.89\% (160/1,007) to 16.58\% (167/1,007), gaining seven correctly solved tasks and 0.70 percentage points. Qwen2.5-Coder-32B-Instruct increases from 16.19\% (163/1,007) to 16.98\% (171/1,007), corresponding to eight additional solved tasks and a 0.79-percentage-point improvement. Similarly, Qwen3.5-27B improves from 19.76\% (199/1,007) to 20.56\% (207/1,007), gaining eight tasks and 0.79 percentage points.

Overall, \textit{SFT} outperforms \textit{Base Contract} for every evaluated model, with absolute improvements ranging from 0.70 to 1.79 percentage points. In relative terms, these improvements correspond to approximately 4.02\%--9.57\% over \textit{Base Contract}. Moreover, for every model, the improvement from \textit{Base Contract} to \textit{SFT} is larger than that from \textit{Vanilla} to \textit{Base Contract}. This pattern indicates that the effectiveness of contract guidance depends not only on providing an intermediate behavioral contract, but also on how that contract is generated. Since the downstream code-generation checkpoint remains unchanged, the observed improvements are associated with specializing the intermediate contract-synthesis stage rather than adapting the code generator itself. 


\mybox{\textbf{Answer to RQ2:}
Contract-specific fine-tuning consistently improves downstream code generation across all five evaluated models while keeping the downstream code-generation model unchanged. Compared with Base Contract guidance, SFT-generated contracts yield relative pass@1 improvements of 4.02\%--9.57\%. Moreover, these gains consistently exceed those obtained by simply introducing Base Contract guidance, indicating that specializing the contract-synthesis stage provides additional downstream benefits.}

\subsection{RQ3: Effectiveness of Contract Validation and Repair}

\paragraph{Approach.}
To investigate whether semantic validation and repair provide additional benefits beyond contract-specific fine-tuning, we compare the \textit{SFT} and \textit{Validated SFT} conditions. Both conditions use the contract-specific SFT model to synthesize the initial behavioral contract and the same original provider-released checkpoint, without the LoRA adapter, for downstream code generation. 

In the \textit{SFT} condition, the synthesized contract is directly provided to the downstream code generator together with the original programming task. In the \textit{Validated SFT} condition, the synthesized contract first undergoes the semantic validation and repair procedure described in Section~\ref{sec:contract-validation}. The validator checks the contract against the original programming task for semantic defects, including incorrect clauses, vague clauses, missing clauses, and expression--description mismatches. When actionable defects are identified, a repair is attempted and the resulting contract is revalidated. The final selected contract may therefore be an accepted repaired contract or the preserved original contract when no repair is accepted.

We first compare aggregate pass@1 between \textit{SFT} and \textit{Validated SFT} on the same 1,007 held-out tasks. To examine the effect beyond aggregate performance, we additionally match the two conditions by task ID and analyze task-level transitions. A \textit{gained} task is one that fails under \textit{SFT} but passes under \textit{Validated SFT}, whereas a \textit{lost} task passes under \textit{SFT} but fails under \textit{Validated SFT}. The difference between gained and lost tasks represents the net change in successfully solved tasks.

\paragraph{Results.}
As shown in Table~\ref{tab:overall_results}, \textit{Validated SFT} achieves the highest aggregate pass@1 among all four experimental conditions for every evaluated model. DeepSeek-Coder-V2-Lite-Instruct improves from 16.68\% (168/1,007) under \textit{SFT} to 17.87\% (180/1,007) under \textit{Validated SFT}, corresponding to 12 additional correctly solved tasks and an absolute improvement of 1.19 percentage points. Qwen2.5-Coder-14B-Instruct increases from 16.58\% (167/1,007) to 17.68\% (178/1,007), gaining 11 tasks and 1.09 percentage points.

Ministral-3-14B-Instruct-2512 improves from 20.46\% (206/1,007) to 21.05\% (212/1,007), corresponding to six additional solved tasks and a gain of 0.60 percentage points. Qwen3.5-27B similarly gains six tasks, increasing from 20.56\% (207/1,007) to 21.15\% (213/1,007), an improvement of 0.60 percentage points. Qwen2.5-Coder-32B-Instruct exhibits the smallest aggregate improvement, increasing from 16.98\% (171/1,007) to 17.38\% (175/1,007), corresponding to four additional solved tasks and 0.40 percentage points. Overall, validation and repair yield absolute improvements of 0.40--1.19 percentage points over using SFT-generated contracts directly.

The task-level analysis in Table~\ref{tab:validation-transitions} provides a more detailed view of these aggregate improvements. For DeepSeek-Coder-V2-Lite-Instruct, 32 tasks that fail under \textit{SFT} become successful after validation, while 20 previously successful tasks become failures, resulting in a net gain of 12 tasks. Ministral-3-14B-Instruct-2512 gains 27 tasks and loses 21, resulting in a net gain of six. Qwen2.5-Coder-14B-Instruct gains 25 and loses 14, producing a net gain of 11. Qwen2.5-Coder-32B-Instruct exhibits substantially fewer transitions, gaining six tasks while losing two. Finally, Qwen3.5-27B gains 22 tasks and loses 16, resulting in a net gain of six.

Inspection of the lost cases reveals examples in which validation or repair
modifies contract guidance that had previously led to a passing
implementation. Such modifications may introduce inaccurate constraints,
remove useful behavioral information, or change the interpretation of
existing requirements. These cases illustrate that, although validation and
repair improve aggregate pass@1, they can also introduce regressions for
individual tasks.

\begin{table}[t!]
    \centering
    \caption{Task-level transitions from SFT to Validated SFT.
    \textbf{Gained} denotes a failure becoming a pass; \textbf{Lost} denotes
    a pass becoming a failure. \textbf{Net gain} is gained minus lost.}
    \label{tab:validation-transitions}
    \small
    \setlength{\tabcolsep}{6pt}
    \renewcommand{\arraystretch}{1.2}
    \begin{tabular*}{\linewidth}{
        @{\extracolsep{\fill}}lrrr@{}
    }
        \toprule
        \textbf{Model}
        & \textbf{Gained}
        & \textbf{Lost}
        & \textbf{Net gain} \\
        \midrule
        DeepSeek-Coder-V2-Lite-Instruct
        & 32 & 20 & 12 \\

        Ministral-3-14B-Instruct-2512
        & 27 & 21 & 6 \\

        Qwen2.5-Coder-14B-Instruct
        & 25 & 14 & 11 \\

        Qwen2.5-Coder-32B-Instruct
        & 6 & 2 & 4 \\

        Qwen3.5-27B
        & 22 & 16 & 6 \\
        \bottomrule
    \end{tabular*}
\end{table}

\mybox{\textbf{Answer to RQ3:}
The combined validation and subsequent code-generation pipeline produces an overall increase in the number of successfully solved tasks. 
Incorporating contract validation and repair (Validated SFT) achieves the highest aggregate code generation performance (pass@1) across all evaluated models.}

\subsection{RQ4: Effect of Contract Quality}

\paragraph{Approach.}
To investigate how the semantic quality of behavioral contracts is associated
with downstream code-generation performance, we analyze the final contracts
used in the \textit{Validated SFT} condition. For each evaluated model, we
consider all 1,007 task--contract records.

Each contract is assessed against its corresponding programming task and
assigned one semantic category: \emph{correct}, \emph{incorrect clause},
\emph{vague clause}, \emph{missing clause}, or
\emph{expression--description mismatch}. Records containing incomplete
contract information that prevents reliable semantic classification are
reported separately and excluded from the five semantic categories.
Two authors independently assessed the contracts against the corresponding
programming tasks, and disagreements were resolved through discussion until
consensus was reached.
When multiple defect types are present, the contract is assigned to the
category corresponding to the defect judged to have the greatest semantic
impact on the contract's representation of the task requirements.

For each semantic-quality category, we associate the contract with the
pass/fail outcome of the corresponding downstream program and compute
category-specific pass@1. This analysis allows us to examine the association
between contract semantic quality and downstream code-generation correctness.

\paragraph{Results.}
\begin{table}[t!]
    \centering
    \caption{Downstream code-generation performance by manually assigned
    contract-quality category in the Validated SFT condition. Each entry reports
    passing programs over contracts in the category, followed by pass@1.}
    \label{tab:manual-contract-quality}
    \scriptsize
    \setlength{\tabcolsep}{3pt}
    \renewcommand{\arraystretch}{1.2}
    \begin{tabular*}{\linewidth}{
        @{\extracolsep{\fill}}lccccc@{}
    }
        \toprule
        \textbf{Model}
        & \textbf{Correct}
        & \textbf{Incorrect}
        & \textbf{Vague}
        & \textbf{Missing}
        & \shortstack{\textbf{Expression--Description}\\\textbf{Mismatch}} \\
        \midrule

        DeepSeek-Coder-V2-Lite-Instruct
        & 64/176 (36.36\%)
        & 59/350 (16.86\%)
        & 7/101 (6.93\%)
        & 47/345 (13.62\%)
        & 3/35 (8.57\%) \\

        Ministral-3-14B-Instruct-2512
        & 119/362 (32.87\%)
        & 5/57 (8.77\%)
        & 33/228 (14.47\%)
        & 46/305 (15.08\%)
        & 9/48 (18.75\%) \\

        Qwen2.5-Coder-14B-Instruct
        & 65/201 (32.34\%)
        & 59/405 (14.57\%)
        & 34/242 (14.05\%)
        & 18/135 (13.33\%)
        & 2/24 (8.33\%) \\

        Qwen2.5-Coder-32B-Instruct
        & 100/395 (25.32\%)
        & 55/357 (15.41\%)
        & 0/12 (0.00\%)
        & 17/198 (8.59\%)
        & 3/40 (7.50\%) \\

        Qwen3.5-27B
        & 131/426 (30.75\%)
        & 34/205 (16.59\%)
        & 3/51 (5.88\%)
        & 42/276 (15.22\%)
        & 2/29 (6.90\%) \\

        \bottomrule
    \end{tabular*}
\end{table}

Table~\ref{tab:manual-contract-quality} reports downstream code-generation
performance according to the manually assigned contract-quality categories.
The distribution of contract quality varies substantially across models.
For DeepSeek-Coder-V2-Lite-Instruct, 176 contracts are classified as correct,
while incorrect and missing clauses account for 350 and 345 contracts,
respectively. Vague clauses occur in 101 cases and
expression--description mismatches in 35 cases. 
Ministral-3-14B-Instruct-2512 produces 362 correct contracts, followed by
305 contracts with missing clauses and 228 with vague clauses. Incorrect
clauses and expression--description mismatches are less frequent, occurring
in 57 and 48 contracts, respectively. Seven records contain incomplete contract information and are therefore
excluded from the five semantic categories. 
For Qwen2.5-Coder-14B-Instruct, incorrect clauses constitute the largest
category with 405 contracts, followed by 242 vague contracts. Correct,
missing, and expression--description mismatch contracts account for 201,
135, and 24 cases, respectively. 

Qwen2.5-Coder-32B-Instruct produces 395 correct contracts and 357 contracts
containing incorrect clauses, together with 198 missing, 12 vague, and
40 expression--description mismatch contracts. Five records contain incomplete
contract information and are therefore excluded from the five semantic
categories.

Finally, Qwen3.5-27B produces 426 correct contracts, 276 contracts with missing
clauses, 205 with incorrect clauses, 51 with vague clauses, and
29 expression--description mismatches. Twenty records contain incomplete
contract information and are therefore excluded from the five semantic
categories.

The downstream results show a consistent association between manually
assessed contract quality and functional correctness. Across all five models,
contracts classified as semantically correct achieve the highest
category-specific pass@1. Correct-contract pass@1 ranges from 25.32\% for
Qwen2.5-Coder-32B-Instruct to 36.36\% for
DeepSeek-Coder-V2-Lite-Instruct. In comparison, contracts containing semantic
defects generally achieve substantially lower downstream success rates. 

\mybox{\textbf{Answer to RQ4:}
Across all five models, contracts classified as semantically correct
consistently achieve the highest downstream pass@1. These results show a
consistent association between contract semantic quality and downstream
functional correctness.}
\section{Discussion}
\label{sec:discussion}

\subsection{Ablation Study on Contract Components}
\label{sec:contract_distribution}

We further investigate whether different behavioral components of a contract
provide distinct forms of guidance for downstream code generation. We focus on
Qwen3.5-27B, which achieves the highest overall pass@1 under the Validated SFT
condition in Table~\ref{tab:overall_results}. For each of the 1,007 evaluation
tasks, we start from the final contract produced by the Validated SFT pipeline
and construct three component-specific variants: \emph{Preconditions Only},
\emph{Postconditions Only}, and \emph{Invariants Only}.

For each variant, we retain only the corresponding behavioral component from
the validated contract and remove the other two components. For example, the
Preconditions Only variant retains all precondition clauses while removing
postconditions and invariants. The original programming task and required
interface information are preserved in every condition. The resulting
component-specific contract is then provided to the same unchanged
Qwen3.5-27B checkpoint used in the main experiment for downstream code
generation.

To ensure a controlled comparison, all other experimental settings remain
identical to those of the Validated SFT condition, including the evaluation
tasks, downstream model, prompt structure, decoding strategy, and execution
procedure. We compare the pass@1 of each component-specific condition with
that obtained using the complete validated contract. This ablation allows us
to examine how much useful guidance can be provided by each behavioral
component in isolation and whether the full contract benefits from combining
multiple complementary forms of behavioral information. 

\begin{table}[t!]
    \centering
    \caption{Component-level ablation of the validated behavioral contract
    for Qwen3.5-27B on the 1,007 held-out TACO tasks. Each entry reports the
    number of passing programs and pass@1.}
    \label{tab:contract-component-ablation}
    \small
    \setlength{\tabcolsep}{8pt}
    \renewcommand{\arraystretch}{1.15}
    \begin{tabular}{lc}
        \toprule
        \textbf{Contract Guidance} & \textbf{Passed / Pass@1} \\
        \midrule
        Preconditions only          & 208 / 1,007 (20.66\%) \\
        Postconditions only         & 211 / 1,007 (20.95\%) \\
        Invariants only             & 204 / 1,007 (20.26\%) \\
        Complete validated contract & 213 / 1,007 (21.15\%) \\ \hline
        Vanilla (no contract) & 198 / 1,007 (19.66\%) \\
        \bottomrule
    \end{tabular}
\end{table}

Table~\ref{tab:contract-component-ablation} presents the component-level
ablation results. All three individual contract components outperform the
Vanilla condition without contract guidance. Vanilla achieves 19.66\% pass@1,
whereas using only preconditions, postconditions, or invariants increases
pass@1 to 20.66\%, 20.95\%, and 20.26\%, respectively. Thus, even an
individual contract component provides useful behavioral guidance beyond the
original programming task alone.

Among the component-only conditions, postconditions achieve the highest
pass@1 at 20.95\%, followed by preconditions at 20.66\% and invariants at
20.26\%. However, no individual component matches the complete validated
contract, which achieves the highest pass@1 of 21.15\%. Compared with
Vanilla, the complete contract improves pass@1 by 1.49 percentage points,
while outperforming the component-only conditions by 0.20--0.89 percentage
points.
These results suggest that individual contract components can provide useful
behavioral guidance. Each component-only condition achieves higher observed
pass@1 than Vanilla, while the complete validated contract achieves the
highest observed pass@1. However, the differences among the component-specific
conditions and the complete contract are relatively small and are therefore
interpreted descriptively.

\subsection{Contract Validation Performance in Repair-Need Detection}
\label{sec:validation_performance}
We further evaluate the contract-validation stage by examining whether it can
correctly identify contracts that require repair.

Under the Validated SFT condition, the five evaluated models each generate
contracts for 1,007 tasks, resulting in 5,035 contract-validation instances.
Since manually analyzing all instances would be time-consuming, we focus on
Qwen3.5-27B, which achieves the highest overall pass@1 under the Validated SFT
condition, and analyze all of its 1,007 contract-validation instances.
For each instance, we record whether the validator accepts the generated
contract or identifies semantic defects requiring repair. Two authors independently inspect each initial SFT-generated contract against
the corresponding programming task to determine whether repair is required.
Disagreements are resolved through discussion until consensus is reached.
This manual assessment serves as the reference for evaluating repair-need
detection.

We formulate repair-need detection as a binary classification problem. A
contract is considered positive if manual inspection identifies at least one
semantic defect requiring repair. We define true positives (TP) as defective
contracts correctly flagged for repair, false positives (FP) as correct
contracts incorrectly flagged for repair, false negatives (FN) as defective
contracts accepted without repair, and true negatives (TN) as correct contracts
accepted without repair. We report precision, recall, F1-score, and accuracy.
Precision is computed as $TP/(TP+FP)$, recall as $TP/(TP+FN)$, F1-score as
$2 \times (\mathrm{Precision} \times \mathrm{Recall}) /
(\mathrm{Precision}+\mathrm{Recall})$, and accuracy as
$(TP+TN)/(TP+FP+FN+TN)$.

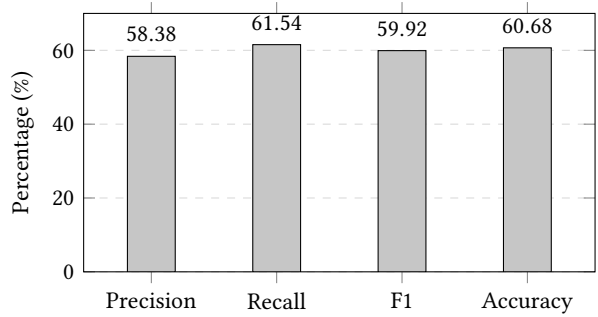
\begin{figure}[t]
\centering
\begin{tikzpicture}
\begin{axis}[
    ybar,
    width=0.6\linewidth,
    height=5cm,
    ymin=0,
    ymax=70,
    ylabel={Percentage (\%)},
    symbolic x coords={Precision,Recall,F1,Accuracy},
    xtick=data,
    bar width=18pt,
    enlarge x limits=0.18,
    ymajorgrids=true,
    grid style={dashed,gray!30},
    nodes near coords,
    nodes near coords style={
        font=\small,
        /pgf/number format/fixed,
        /pgf/number format/precision=2
    },
    every node near coord/.append style={yshift=2pt},
    tick label style={font=\small},
    label style={font=\small},
]

\addplot[fill=gray!45, draw=black]
coordinates {
    (Precision,58.38)
    (Recall,61.54)
    (F1,59.92)
    (Accuracy,60.68)
};

\end{axis}
\end{tikzpicture}

\caption{Repair-need detection performance of the contract validator
for Qwen3.5-27B.}
\label{fig:validator-performance}
\end{figure}
Figure~\ref{fig:validator-performance} shows the performance. As we can see, the validator produces 296 true positives, 211 false
positives, 185 false negatives, and 315 true negatives. It therefore achieves
58.38\% precision, 61.54\% recall, 59.92\% F1-score, and 60.68\% accuracy, demonstrating that automated semantic validation can identify
many repair-worthy contracts, but remains far from perfect. Both false
positives and false negatives are substantial, motivating more reliable
contract-validation techniques. 

\section{Threats to Validity}
\label{sec:threats}

\textbf{Construct Validity.}
We measure functional correctness using pass@1 on the executable TACO test
suites, where a program is correct only if it passes all available tests.
However, these tests may not cover every behavior specified by a task, and
passing them does not constitute a formal proof of correctness. Similarly,
our contract validator detects defects according to four predefined semantic
categories; thus, a \texttt{CORRECT} decision indicates that no such defect
was detected rather than formal verification of the contract. Moreover, the
same SFT model performs contract generation, validation, and repair, which may
introduce correlated errors. The four defect categories are also not
exhaustive and may omit properties such as efficiency or security constraints.

\textbf{Internal Validity.}
We isolate changes to the contract pipeline while keeping the downstream
code-generation model and settings fixed within each model family. We use
greedy decoding with one candidate per task to eliminate sampling variation.
Nevertheless, results may depend on prompts and implementation choices.
Our SFT corpus undergoes automatic filtering, manual review, cleaning, and
deduplication, which may introduce selection or human-judgment bias.
Evaluation tasks are held out from SFT to reduce direct data leakage.

The RQ4 analysis is observational: task characteristics such as difficulty
may affect both the likelihood of generating a semantically correct contract
and the likelihood of generating a correct program. The category-specific
pass@1 differences should therefore be interpreted as associations rather
than causal effects of contract quality.

\textbf{External Validity.}
We evaluate TACO tasks from four algorithmic categories and three primary
code-model families. While this provides diversity in tasks and models, our
findings may not generalize to repository-level development, API-intensive or
interactive applications, other programming domains, substantially different
model architectures, proprietary models, or future code-generation systems.

\textbf{Conclusion Validity.}
All conditions are evaluated on the same held-out tasks, enabling paired
task-level comparisons. However, some of the observed pass@1 differences are
small and should therefore be interpreted cautiously. Furthermore, each
condition generates one deterministic candidate per task; our conclusions
therefore concern single-generation pass@1 and do not capture pass@k or
variation under stochastic generation.
\section{Conclusion}
\label{sec:conclusion}

This paper presents CG$^{2}$, a contract-guided framework for improving LLM code generation by introducing behavioral contracts as structured intermediate specifications between natural-language programming tasks and generated code. CG$^{2}$ specializes contract synthesis through parameter-efficient fine-tuning and further applies semantic validation and repair to improve contract quality, while keeping the downstream code-generation model unchanged.

We evaluate CG$^{2}$ on 1,007 held-out TACO tasks across five code-oriented LLMs. The results consistently show that introducing behavioral contracts improves pass@1 over direct generation, contract-specific fine-tuning provides additional gains, and semantic validation and repair further improve aggregate performance. Overall, CG$^{2}$ improves pass@1 over direct code generation by 1.49--3.27 percentage points, corresponding to relative improvements of 7.58\%--22.45\%, and over contract guidance without fine-tuning by 1.19--2.38 percentage points, corresponding to relative improvements of 7.04\%--14.65\%. Our contract-quality analysis further identifies recurring semantic defects, including incorrect, vague, and missing clauses and expression--description mismatches. Across all five models, contracts classified as semantically correct are consistently associated with substantially higher downstream pass@1 than defective contracts.

These findings provide evidence for separating behavioral specification from implementation in LLM-based code generation. Improving the generation and semantic quality of intermediate behavioral contracts can provide more reliable guidance without modifying the downstream code generator. Future work can extend this approach to larger models, broader software-engineering tasks, richer contract representations, and stronger verification and repair techniques.

\section*{Data Availability}  
A replication package containing our data and source code is publicly available at
\url{https://anonymous.4open.science/r/Contract-Guided-LLM-Code-Generation-B9FB/}.

\bibliographystyle{ACM-Reference-Format}
\bibliography{references}

\end{document}